\documentclass[11pt,a4paper]{article}

\usepackage{a4wide}
\usepackage{authblk}
\usepackage{graphicx}
\usepackage{xspace}

\graphicspath{{./figures/}} 

\newcommand{\etal}{\emph{et~al.}\xspace}

\title{Designing Problem Sessions for Algorithmic Subjects to Boost Student Confidence}

\author[1]{Andr\'e van Renssen}
\affil[1]{The University of Sydney, Australia}

\date{}

\begin{document}
\maketitle

\begin{abstract}
In this paper, we describe how we changed the structure of problem sessions in an algorithmic subject, in order to improve student confidence. The subject in question is taught to very large cohorts of (around 900) students, though our approach can be applied more broadly. We reflect on our experiences over a number of years, including during the pandemic, and show that by adding clear sectioning indicating the style of the questions and by including simple warm-up questions, student engagement and confidence improves, while making the teaching activities of our teaching assistants easier to manage. 
\end{abstract}

\section{Introduction}
Algorithmic subjects are taught as part of every Computer Science curriculum~\cite{cc2020}. Often a curriculum contains multiple algorithmic subjects, focusing on different topics in increasing difficulty. For example, the first algorithmic subject students encounter might focus on explaining basic data structures and simple graph algorithms, while later subjects could cover dynamic programming, flow networks, and complexity theory. The key outcomes of these classes are two-fold. The first is to introduce a number of standard concepts and techniques to the students to ensure they can recognize and use these to solve common problems, such as sorting and graph traversal. Second, students should take this knowledge and be able to use it to design new algorithms and data structures for new problems, thus allowing them to grow their knowledge and apply it in their future subjects and careers. 

One way algorithmic subjects are often delivered, is with a combination of lectures, 
which explain the main ideas and show examples, and problem sessions, where students practice their analysis and algorithm design skills with a teaching assistant to provide hints and feedback. The problem sessions are generally the first (supervised) moment where students can try their hand at exploring the algorithms they were shown in the lectures as well as at designing algorithms for different problems themselves. However, making the step from being shown an algorithm and seeing its analysis to applying the skills to similar problems themselves can be a daunting one. Hence, problem sessions need to inspire confidence, so that students remain engaged. 

Problem sets used by the previous instructor for this subject consisted of a set of questions (where the students analysed variant algorithms, or adapted algorithms to new situations, etc.) ordered (roughly) in order of increasing difficulty. While this indeed allowed the students to practice designing algorithms themselves, we got the feedback that the students found the step between the lecture and the problem sessions rather large and as a result they sometimes felt lost trying to follow along during the problem sessions, making it harder for students to participate. We note that this was not reflected in the student grades, but rather in the student level of confidence in their own abilities. 

In this paper, we discuss the approach we took in re-designing the problem sessions to make the step from the lectures to the problem sessions easier to bridge for the students, aiming to increase student confidence and engagement. We re-designed our problem sessions in a few seemingly basic manners that can be applied to a wide range of subjects. Specifically, we added two things: section headers, which give a rough indication of the style of the questions in that section and thus help them understand better what is expected of them, and warm-up questions that allow the students to first confirm their understanding of the lecture material before being asked to apply it to harder problems. The latter is intended to make the step from the lectures to the problem sessions easier to bridge and typically asks the students to trace an algorithm from the lectures on a problem instance to ensure they understand the algorithms in sufficient detail to use them to design more complicated algorithms in later sections of the problem session. 

We reflect on the changes we introduced. From an instructor point of view, we hoped that especially the new warm-up questions would bridge the perceived gap between the lectures and the problem sessions. We focused on the learning experience, hoping that by re-designing the problem sessions the students would find the material more accessible, thus allowing students to feel more engaged during the problem sessions and increasing their confidence in their own algorithmic problem solving abilities. We observed that students were indeed more engaged throughout the semester. Student comments, obtained through the End of Semester Survey, also mentioned how the problem sessions helped them confirm their knowledge before applying it to more difficult scenarios. 

As well as looking at student reports, we also asked the teaching assistants about their experiences. They were very positive, noticing changes in student understanding, engagement, and in their own experience running the problem sessions. We did this both to confirm student statements, as well as to obtain comparative data over a number of years, as a number of teaching assistants have worked with us to deliver this subject on multiple occasions.

\section{Related Work}
\label{sec:related}
Teaching algorithmic subjects has received significant attention in the literature, ranging from studies addressing what an algorithmic subject and its topics actually are~\cite{hertz2010cs1,luu2023algorithms} to designing alternative grading methods for non-major student cohorts~\cite{weber2023using}. Luu \etal~\cite{luu2023algorithms} performed a large scale survey on the topics and techniques covered in algorithmic subjects across the United States. A smaller scale study was carried out by Hertz~\cite{hertz2010cs1}, who focused on the content of CS1 and CS2 subjects. Additionally, there has been work on identifying common misconceptions of students on the topics of algorithmic subjects~\cite{farghally2017towards,ozdener2008comparison,shindler2022student,velazquez2019students,zehra2018student}. 

Unfortunately, related work on improving student engagement and confidence in problem sessions is very scarce. Some work on improving student motivation and engagement has focused on adapting the grading method to facilitate this~\cite{spurlock2023improving}. Spurlock~\cite{spurlock2023improving} implemented the \emph{ungrading} technique, eliminating numeric grades, allowing the resubmission of assignments, and encouraging student input to their final grade. Using this approach, students commented on reduced anxiety and increased control over their subject outcome. Additionally, they report that instructor-student conversations focused more on the assignment feedback than on the assignment grade. Our institution's central assessment policies, and student expectations from all their other classes, would make this change difficult to introduce. 

Significant work has been done in the area of active learning (see Brame~\cite{brame2016active} for a recent overview). The seminal work by Bonwell and Eison~\cite{bonwell1991active} discusses a great number of techniques. While some of these techniques can be applied in our setting, the majority is infeasible due to low student attendance, being a commuter university and all lectures being available in recorded form by university policy. McConnell~\cite{mcconnell1996active} also discusses different active learning methods, including methods such as tracing algorithms and physical activities to illustrate for example a token passing algorithm. In our setting, physical activities are infeasible due to this method not reaching the majority of students. Tracing algorithms is indeed a method we can use in our lectures (and indeed do), however, this method did not achieve the student confidence we aimed for. Schweitzer and Brown~\cite{schweitzer2007} described the use of visualisations in lectures to facilitate student engagement, including their use to explain algorithms. We apply this method as well, but it did not achieve the intended level of student confidence. 

Unfortunately, as far as we are aware, none of the literature considers student confidence in their own abilities as the main driver for change and most literature focuses on the lecture rather than the problem sessions.

\section{Educational Setting}
\label{sec:setting}
While problem sessions can be designed using our approach for a variety of subjects, in this paper we describe our experiences in teaching the algorithmic subject COMP2123. This is a sophomore-level subject on Data Structures and Algorithms, taught at a large state university. It covers many of the topics traditional in a CS2 class, including big \emph{O} notation and running time and space analysis; the standard list, tree, dictionary, and graph data structures and their algorithms for searching, insertion, deletion and traversal. It also revises recursion and sorting, and introduces some basic algorithm design techniques such as greedy and divide-and-conquer. In our curriculum, this is taken in a student's third semester, following two semesters in which they learn programming first imperatively in Python and then object-oriented in Java. The subject is core in the Computer Science major, but it is also taken by students from other fields such as Electrical Engineering. Due to institutional policy, we cannot enforce other prerequisites, and so we do not assume prior learning of discrete math. The subject has grown in recent years from under 500 to around 900 students. The lectures were delivered online, but most of the students attended in-person problem sessions in the most recent delivery. 

\subsection{Subject Structure}
\label{sec:structure}
The subject is structured as follows: Every week of the 13-week semester there is a 2-hour lecture for the whole student cohort and a 2-hour problem session for every 20-30 students with one teaching assistant (usually a senior undergraduate or PhD student). The lecture includes active learning elements, for example by means of algorithm tracing and visualisations and short questions, such as "What is the degree of this vertex?" or "Which of these greedy algorithms will give the optimal solution in all instances?". We note that the effectiveness of these elements is limited and was deemed insufficient on its own, due to the institutional culture of low lecture attendance -- being a commuter school, with default recording of all lectures.  

The number of students taking this subject doubled over the years related to this experience report, where in the final two years there were on average 40 problem sessions run by about 20 teaching assistants. During these problem sessions, the students are encouraged to collaborate in groups of 3-4 students to solve a series of problems, provided to them a week before the problem session to allow them to prepare. The teaching assistant is there to help the students when they get stuck and to go through some problems with all attendees. In the first part of the problem session this is intended to give the students some ideas on how to approach the posed problems, while later it is mostly used to help students with the questions that a large number of them struggled with. The answers to all problem session questions are released after the last problem session on that topic finishes. Note that the students' attendance and work in problem sessions, and their answers to the problems, do not contribute to their final grade. Instead, they are formative, to build skills which are later assessed in homework assignments and in the final exam.

In order to ensure that all students have access to the lecture, also when they fall ill or are otherwise unable to attend, all lectures are recorded and made available to the students through the online learning management system. This system also provides a discussion board where the instructor and the teaching assistants help answer any questions the students may have about the subject and its content. We encourage students to also answer each other's questions, as this helps them learn to explain their ideas and the discussion board enables us to endorse answers to assure the student asking the question that it was answered correctly. 

To ensure that students are keeping up with the material and to give them early feedback on their understanding, there are 10 short 15-minute quizzes that the students complete at home, each worth 1\% of their final grade. Each quiz consists of between 6 and 10 questions, testing fundamental components of the lecture material in the form of multiple choice questions. Questions range from true/false statements to asking which edges of a weighted graph would be part of its minimum spanning tree. 

To test their algorithmic problem solving skills, there are 5 fortnightly homework assignments, each worth 6\% of their final grade. These assignments each contain three questions: one relatively easy question where they are asked to trace an algorithm from the lectures on a given example, followed by two questions where they are asked to design an efficient algorithm for a given problem. 

Finally, at the end of the semester there is a written exam worth 60\% of the student's final grade. This high weight is strongly encouraged by the university to limit the impact of potential cheating on the assessments the students do at home and thus ensure that students indeed acquire the majority of their grade under controlled circumstances. To pass the subject, students need to obtain at least 50\% on the subject and score at least 40 out of 100 on the final exam. While this exam barrier may seem harsh, the exam is designed in such a way that it can be met by performing basic running time analysis and tracing algorithms they saw in the lectures. 

The subject comes in two flavors: the regular stream and the advanced stream. The regular stream is taken by most students, while the advanced stream is intended for the high-achieving students (about 10\% of the class). The structure described above applies to both streams and the main difference between the two is that students in the advanced stream are additionally taught some more complicated aspects of the topics. This can, for example, include amortized analysis, in-depth proofs of expected running times, and the divide-and-conquer algorithm for convolutions. They are also challenged with a few additional hard questions in their problem sessions and slightly harder assignment questions.

\section{Problem Sessions for Data Structures and Algorithms}
\label{sec:main}

\subsection{Previous Approach}
The previous problem sets were designed by an experienced instructor to use the limited problem session time to give students the most new experiences rather than revision. The problem sets were simply a list of 5-10 problems in roughly increasing level of difficulty. These questions would assume that students are familiar with the lecture content and thus focused solely on applying this knowledge to new problems. As an example, we present the first problem of the problem set on hashing (see Problem~1) and the first two problems of the problem set on graph algorithms (see Problems~2 and 3): 

\vspace{0.5em}

\noindent \textbf{Problem 1.} Design a sorted hash table data structure that performs the usual operations of a hash table with the additional requirement that when we iterate over the items, we do so in the order they were inserted into the hash table. Iterating over the items should take $O(n)$ time where $n$ is the number of items stored in the hash table. Your data structure should only add $O(1)$ time to the standard put, get, and delete operations.

\vspace{0.5em}

\noindent \textbf{Problem 2.} Let $G=(V,E)$ be an undirected graph with edge weights $w:E \rightarrow R^+$. For all $e \in E$, define $w_1(e) = \alpha w(e)$ for some $\alpha > 0$, $w_2(e) = w(e) + \beta$ for some $\beta > 0$, and $w_3(e) = w(e)^2$.
\begin{enumerate}
  \item[(a)] Suppose $p$ is a shortest $s$-$t$ path for the weights $w$. Is $p$ still optimal under $w_1$? What about under $w_2$? What about under $w_3$?
  \item[(b)] Suppose $T$ is a minimum weight spanning tree for the weights $w$. Is $T$ still optimal under $w_1$? What about under $w_2$? What about under $w_3$? 
\end{enumerate}

\noindent \textbf{Problem 3.} It is not uncommon for a given optimization problem to have multiple optimal solutions. For example, in an instance of the shortest $s$-$t$ path problem, there could be multiple shortest paths connecting $s$ and $t$. In such situations, it may be desirable to break ties in favor of a path that uses the fewest edges.

Show how to reduce this problem to a standard shortest path question. You can assume that the edge lengths $\ell$ are positive integers.
\begin{enumerate}
  \item[(a)] Let us define a new edge function $\ell'(e) = M \ell(e)$ for each edge $e$. Show that if $P$ and $Q$ are two $s$-$t$ paths such that $\ell(P) < \ell(Q)$ then $\ell'(Q) - \ell'(P) \geq M$.
  \item[(b)] Let us define a new edge function $\ell''(e) = M \ell(e) + 1$ for each edge $e$. Show that if $P$ and $Q$ are two $s$-$t$ paths such that $\ell(P) = \ell(Q)$ but $P$ uses fewer edges than $Q$ then $\ell''(P) < \ell''(Q)$.
  \item[(c)] Show how to set $M$ in the second function so that the shortest $s$-$t$ path under $\ell''$ is also shortest under $\ell$ and uses the fewest edges among all such shortest paths. 
\end{enumerate}

\vspace{0.5em}

In terms of the revised Bloom's taxonomy~\cite{krathwohl2002bloom} these questions map to the \emph{Apply} and \emph{Analyze} categories. The students experienced this as a large jump from what they knew how to do, and a number of students were disheartened when they could not immediately solve at least some of the problems. Some example comments are shown below (terminology updated for consistency with standard terms): 

\begin{itemize}
  \item I personally felt that the difficulty level of the lecture compared to the problem sets/as-signments was huge; the lecture being significantly easier than the content of the problem sessions/assignment. 
  \item I felt the problem sets were very difficult and frustrating. 
  \item Problem sets are not organised in order of difficulty making it hard to build understanding and confidence. Some really basic questions with worked solutions at the start of the problem sets would be useful.
  \item The disparity between the difficulty of the lecture content and the problem sets made it difficult to apply knowledge learnt from lectures to problem sets.
  \item The problem sets were often way beyond my own and some classmates ability to complete. 
  \item Problem session content always felt like quite a big step up in difficulty.
  \item The difficulty of problem sets was very daunting and intimidating considering what we had just learnt in the lecture which was very general. Rather than throw us into the deep end with challenging questions from the get go, there need to be some intermediate questions that will prepare us for tackling the tricky questions.
\end{itemize}

The teaching assistants also struggled with the fact that they were trying to help the students solve questions while at the same time trying to identify misconceptions about the lecture material. The latter was especially challenging, as it was hard to identify whether an issue arose from a misconception in the lecture material or because the student did not understand the question.  

\subsection{New Approach}
When the author first taught this subject in 2020, we were made aware of the student comments by the previous, experienced, instructor. To reduce the gap between the lectures and the problem sessions and to give the students more confidence in their own abilities, we restructured the problem sessions. First, we split the problem set into three different sections to give the students an indication of the style of the problems in each section. The sections we used are titled \emph{Warm-up}, \emph{Problem solving}, and \emph{Advanced problem solving}. 

The \emph{Warm-up} section is used to make the step from the lecture to the problem session smaller and typically contains questions that ask the students to trace an algorithm they had been taught, coming up with worst-case examples for an algorithm from the lectures, or proving basic properties of data structures and algorithms which they had been taught already. The questions are intentionally on the easy side, to give the students confidence in their understanding of the material, while ensuring that misconceptions are identified early. While no explicit time limit was given to the teaching assistants for these questions, in most cases these questions took 15-30 minutes to go through. In terms of the revised Bloom's taxonomy~\cite{krathwohl2002bloom}, these questions map to \emph{Remember}, \emph{Understand}, and \emph{Apply}. What is used from the student's learning is the most directly presented material, such as an algorithm or analysis result. As essentially none of these existed in the problem sets, we designed new ones for each problem session. 

The \emph{Problem solving} section largely maps to the existing questions and we did not make any major changes here. These questions are the main focus of the problem sessions, to ensure that the students' understanding and skills to apply their knowledge to new problems grows to the level needed to successfully complete the assignments. These questions typically map to \emph{Apply} and \emph{Analyze} in the revised Bloom's taxonomy~\cite{krathwohl2002bloom}. What is used from the student's learning is more abstract or meta-level material, such as an algorithm design technique or an analysis technique.

Finally, the \emph{Advanced problem solving} section goes beyond what the students are expected to be able to do for the assignments and is mostly intended for the high-achieving students in the advanced stream of the subject. These questions typically map to the \emph{Evaluate} and \emph{Create} in the revised Bloom's taxonomy~\cite{krathwohl2002bloom}. These questions can be simplified research questions (where the answers are known) or follow-up problems related to some of the advanced content covered in the lectures. Some of these questions were part of the existing problem sessions, but these questions are regularly updated as they tend to come from the research topics of the instructor. 

Typically a problem set contains 2-3 questions in the \emph{Warm-up} section, 5-7  questions on \emph{Problem solving}, and 2-3 challenging questions in \emph{Advanced problem solving}. Students are encouraged to complete the warm-up questions before the problem session as much as possible. A problem session starts with checking if everyone was able to do these warm-up questions and if not, briefly go through the ones that students struggled with, as this generally indicates that a topic from the lectures was not quite clear to the students. This adds a learning moment to the problem sessions early on to catch common misconceptions about the material. Some students also attempt the problem solving questions on their own, but these questions remain the main focus of the problem session and teaching assistants are instructed to aim to complete at least two such problems during the problem session.

\subsection{Example Problem Set}
This section contains some example questions from the three types of problems in our new problem sessions (renumbered for this paper). We note that compared to the previous problem sets, Problems~1-3 did not exist. Problems like Problems~4-6 existed, though Problem~5 was added as a slightly easier problem solving question than the existing Problem~6. Finally, Problems~7 and~8 are new, though advanced questions (including Problem~9) existed in most other problem sessions. 

Below, we show three questions that are part of the \emph{Warm-up} section to test the students' understanding of some of the algorithms discussed during the lecture. The first is from the problem session on hashing (see Problem~1), while the remaining two are used in the problem session on graph algorithms, focusing on Dijkstra's shortest path algorithm (Problem~2) and both Prim's and Kruskal's approach to constructing a minimum spanning tree (Problem~3). 

\vspace{0.5em}

\noindent \textbf{Problem 1.} Consider a hash table of size $N > 1$, and the hash function such that $h(k)= k \bmod 2$ for every $k$. We insert a dataset $S$ of size $n < N$. After that, what is the typical running time of $\textsc{get}$ for chaining and open addressing (as a function of $n$)?

\vspace{0.5em}

\noindent \textbf{Problem 2.} Consider Dijkstra's shortest path algorithm for undirected graphs. What changes (if any) do we need to make to this algorithm for it to work for directed graphs and maintain its running time? 

\vspace{0.5em}

\begin{figure}[ht]
 \centering
  \includegraphics[scale=0.35]{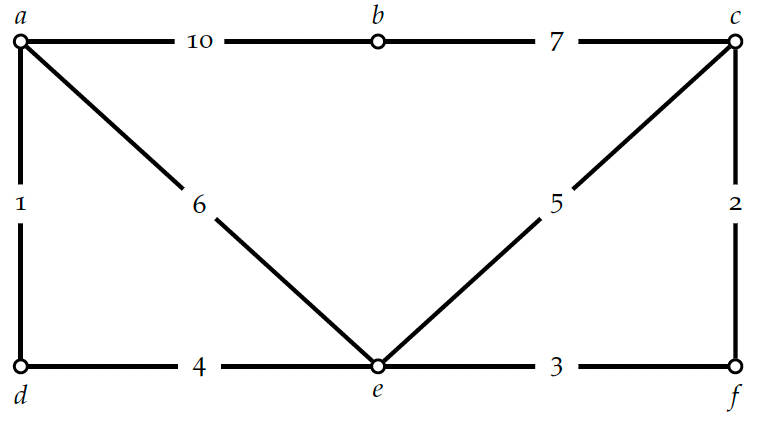} 
  \caption{A weighted graph.}
  \label{fig:graph}
\end{figure}

\noindent \textbf{Problem 3.} Consider the weighted undirected graph $G$ shown in Figure~\ref{fig:graph}. 

Your task is to compute a minimum weight spanning tree $T$ of $G$:
\begin{enumerate}
  \item[(a)] Which edges are part of the MST? 
  \item[(b)] In which order does Kruskal's algorithm add these edges to the solution? 
  \item[(c)] In which order does Prim's algorithm (starting from $a$) add these edges to the solution? 
\end{enumerate} 

\vspace{0.5em}

Examples of questions that fall in the \emph{Problem solving} category can be seen below. Problem~4 comes from the problem session on binary search trees and Problems~5 and 6 are from the problem session on greedy algorithms and give the students the opportunity to design such algorithms for problems they have not seen before. Though not explicitly stated, each of the problems asks the students to design the algorithm, then prove its correctness, and finally analyze its running time. 

\vspace{0.5em}

\noindent \textbf{Problem 4.} Consider the following operation on a binary search tree: {\sc second-largest}$()$ that returns the second largest key in the tree. Give an implementation that runs in $O(h)$ time, where $h$ is the height of the tree.

\vspace{0.5em}

\begin{figure}[ht]
\centering
\includegraphics{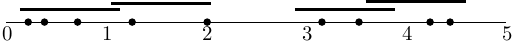} 
\caption{Covering points with unit intervals.}
\label{fig:Covering}
\end{figure}

\noindent \textbf{Problem 5.} Design a greedy algorithm for the following problem (see Figure~\ref{fig:Covering}): Given a set of $n$ points $\{x_1, ..., x_n\}$ on the real line, determine the smallest set of unit-length intervals that contains all points. 

\noindent \textbf{Problem 6.} Suppose we are to schedule print jobs on a printer. Each job $j$ has an associated weight $w_j > 0$ (representing how important the job is) and a processing time $t_j$ (representing how long the job takes). A schedule $\sigma$ is an ordering of the jobs that tells the printer in which order to process the jobs. Let $C_j^\sigma$ be the completion time of job $j$ under the schedule $\sigma$.

Design a greedy algorithm that computes a schedule $\sigma$ minimizing the sum of weighted completion times, that is, minimizing $\sum_{j} w_j C_j^\sigma$. 

\vspace{0.5em}

Finally, we highlight some of the advanced problems, in this case related to the advanced research-related topic of routing on geometric graphs (Problems~7 and 8) and the divide-and-conquer algorithm for convolutions (Problem~9). These topics were covered only for the students in the advanced stream. 

\vspace{0.5em}

\noindent \textbf{Problem 7.} Show that greedy routing\footnote{Greedy routing is a routing algorithm that forwards the message by sending it to the neighbor of the current vertex that is closest to the intended destination.} can still get stuck even if it remembers all previously visited vertices and ignores those when determining the next vertex on the path. 

\vspace{0.5em}

\noindent \textbf{Problem 8.} Show that compass routing\footnote{Compass routing is a routing algorithm that forwards the message by sending it to the neighbor $v$ of the current vertex $u$, such that the angle between edge $uv$ and the line connecting $u$ to the intended destination is minimized.} doesn't always reach the destination in general graphs. For an extra challenge, try to construct an example where it cycles through a set of vertices larger than 3. 

\vspace{0.5em}

\noindent \textbf{Problem 9.} In class we saw an $O(n \log n)$ time algorithm for computing the convolution of two vectors\footnote{The convolution of two vectors corresponds algebraically to the multiplication of two polynomials whose coefficients are defined by the elements of the vectors.} of length $n$. The convolution operator can be defined for vectors of unequal length, say $n$ and $m$ where $m< n$. Design an algorithm for this problem that runs in $O(n \log m)$ time.

\section{Reflection}
\label{sec:reflection}
In terms of the overall student grades, the effect of the changes we made to the problem sessions seem minor. While the number of students not showing up for the exam decreased by a few percent and the average student grade as well as the maximum student grade increased by a few points, we cannot conclude that this was because of the changes made to the problem sessions, as these numbers fluctuate a bit from one year to another. 

\begin{table*}[tb]
\small
\begin{center}
\begin{tabular}{ |c|c||c|c|c| } 
 \hline
  & 2019 & 2020$^*$ & 2021$^*$ & 2022$^*$ \\ 
  \hline
 Number of students & 481 & 545 & 889 & 894 \\ 
 Response rate & 25\% & 51\% & 37\% & 31\% \\ 
 \hline
 Overall mean rating over all questions & 3.87 & 4.07 & 4.17 & 4.16 \\
 \hline
 Results from 
 related to the problem sessions: & & & & \\
 1. The work has been intellectually rewarding. & 4.02 & 4.23 & 4.26 & 4.29 \\
 2. I developed relevant critical and analytical thinking skills. & 4.16 & 4.27 & 4.30 & 4.32 \\
 3. I have had good access to valuable learning resources. & 3.86 & 4.09 & 4.14 & 4.17 \\
 4. The problem sessions effectively supported & 3.91 & 3.80 & 4.14 & 4.08 \\
 my learning and were worthwhile. & & & & \\
 \hline
\end{tabular}
\end{center}
\caption{Overview of the End of Semester Survey results. 2019 was taught by the previous instructor and the year before the changes to the problem sessions were made. 2020-2022 all had the new problem session structure, with minor refinements made to the questions over the years. Years marked with $^*$ were heavily impacted by the pandemic. Scores for the various questions are out of 5, where higher scores indicate better outcomes.}
\label{table:uss}
\end{table*}

There were, however, significant changes to the End of Semester Survey outcomes related to this subject. This survey is run institution-wide every semester for every subject and all students are asked to fill it out. In large subjects such as this one, the response rate tends to be around 30\%. Students are asked about a variety of aspects of the subjects, including whether the work they did was intellectually rewarding, whether they had good access to valuable learning resources, and whether the problem sessions effectively supported their learning. Questions use a five-point Likert scale. Student ratings are aggregated, comments are anonymous, and the survey responses are released to the instructor only after the final exam grades are finalized. An overview of the results related to the problem sessions can be seen in Table~\ref{table:uss}. 

We observe that in the year 2020, when the newly structured problem sessions were introduced, the students' responses found the work more intellectually rewarding, better for developing their critical and analytical thinking skills and they found that they had better access to valuable learning resources (all scores increased by 0.11-0.25 out of 5 with respect to 2019). By making minor modifications to the problem sessions, we refined these outcomes even further for the 2021 and 2022 sessions of this subject. 

In terms of whether the problem sessions effectively supported their learning, the cohort of students in 2020 rated this lower than the year before (3.80 out of 5, down by 0.11 with respect to the previous year). This is likely due to the fact that this subject is taught in the February-June semester, which coincided with the start of the pandemic and from one week to the next all lectures and problem sessions were moved from in person to remote. This switch along with the fact that none of the teaching assistants had taught online before are likely reasons for this lower than expected rating. When we got used to teaching online, the students rated this same point significantly higher (4.14 out of 5, up by 0.34 with respect to the previous year). 

\subsection{Student Comments}
Next, we look more closely at the student comments pertaining to the problem sessions. These comments also come from the End of Semester Surveys, so students can comment on what they found positive about the subject and what they feel should be improved for the next year. 

Below we list a selection of the student comments pertaining to the problem sessions (terminology updated for consistency with standard terms): 
\begin{itemize}
\item This subject makes [me] feel much more self-confident about my capacity and intelligence. Now I know [basic] algorithms is something everyone can learn and be good at. It is not for only smart people as the stereotype.
\item I really enjoyed how the lectures and problem sessions were melded together through the first few warm-up questions in each problem session so that we would have a chance to utilise the lecture content directly before jumping into tougher questions.
\item Questions were not too difficult or too simple to do, was a nice challenge. 
\item The problem sessions have been very helpful and really helped breakdown how to approach questions. 
\item Problem sessions were really good at backing up what was taught during class.
\item ...problem session content was good reinforcement.
\item Problem session material was well structured to learn more about the algorithms that we learned in the lecture.
\item ...the problem sessions effectively reinforced content in the lectures. 
\item ...the problem session problems have been really enlightening.
\item ...the problem sessions were always clearly applicable to the content. Even though the problem sets were challenging, seeing them in the problem session taught us how to solve difficult questions for our assignments.
\item Though challenging, [...] the problem sessions do a great job of leading us through step by step to get to a competent level of completion by the end of the subject. This felt especially important considering this is quite unlike anything we have completed before. 
\end{itemize}

As can be seen from the students' responses, they felt that the problem sessions were structured in a way that helped them learn by easing them into the problem sets using the warm-up questions and gradually increasing the level of difficulty, as intended. They also found the questions initially challenging, but very helpful in their learning process. Hence, from the student perspective the new problem sets had the exact effect we were hoping for. 

\subsection{Teaching Assistant Comments}
Finally, we look at the impact the new, more structured problem sessions had from the teaching assistant perspective. These comments come from an open enquiry sent to the teaching assistants after the subject had concluded. They were encouraged to write about any changes they noticed compared to their previous experience in the subject. 

Below we list a selection of their responses (terminology updated for consistency with standard terms): 
\begin{itemize}
  \item ...the warm-up questions are helpful for students to remind them about the lecture contents 
  [...]
  Compared to before we had them, there would be at least half the class who would just have no idea about the main problem solving questions unless given an opportunity to review the lecture material, and the warm-up questions helped them do this and ease them into it. It improved their learning process by motivating them to take initiative with their learning since the warm-up questions were easier and directly related to lecture content, helping them develop a link to be more able to solve the harder questions and also giving them more of a sense of satisfaction at the end.
  \item 
  From my experience teaching in person this year with very similar problem sessions, the warm-up questions often are good to identify basic flaws in their understanding of lecture content, which is helpful for the later questions as I can generally focus just on how to solve the problems rather than simultaneously tackle that and basic definitions.
  \item From my experience, the warm-up questions have been well-received by the majority of the students in the class. In fact, more than 90\% of all my students (90 students in total) have found them to be useful in preparing for the problem sessions. 

  The warm-up questions serve as an excellent opportunity for students to review the material covered in class before or at the start of the problem session. By attempting the questions ahead of time, they are able to build confidence in their understanding of the concepts and identify any areas where they may need additional clarification.

  Moreover, the warm-up questions help to create a more engaging learning experience for students. They are more focused on attempting the remaining questions as a result of their prior review, confidence and familiarity with the material.

  Additionally, the warm-up questions also enable the teaching assistant to assess the students' level of understanding before the problem session begins. This allows the teaching assistant to tailor their instruction and provide more targeted support to students who may be struggling with certain concepts.

  Overall, the warm-up questions have been a valuable addition to the problem sets, and I believe they have contributed to the improved engagement and understanding of the students in the class.
  \item The students of mine that did the warm-ups found them very useful. I would say that basically all of them came into the problem sessions not needing assistance for the majority of the other questions. 
  \item While they definitely found [the warm-up questions] pretty easy (especially compared to the rest of the problem set), it helped solidify the foundations of the topic that we were focusing on in that week, and helped clarify any confusion or questions that the students had before we launched into the rest of the problem session. While I'm not sure if the students noticed, I definitely found that I was answering questions/clarifying in these warm-up questions that were necessary for future questions. 
  \item I definitely feel the advanced sections were sufficiently challenging. Even running advanced classes,
  we were never at risk of running out of content, and it took us time to step through the advanced questions, often requiring input from different students to solve, which I loved.
\end{itemize}

From the above comments, we can see that the more structured problem sets allowed them to address any misconceptions the students had about the material before starting the actual problem solving questions, making running their problem sessions easier and more effective. Since the warm-up questions are a bit easier, they also noticed that solving these created a more engaging environment for the students, as the students were on a more similar level of understanding, which in turn led to better understanding, higher satisfaction, and better student confidence in their abilities. 

Hence, while the students themselves may not have actively noticed the difference, the teaching assistants teaching them in the problem sessions noticed a significant difference in understanding of their student cohort.

\subsection{Limitations on Our Reflections}
\label{sec:limitations}
There are a few factors that could affect the above discussion. First, there is the fact that the new problems sets were introduced in the year that teaching also moved to become fully online due to the pandemic. However, we believe that if this had an effect, we would expect this to be a negative one, as online teaching was generally regarded more negatively than in person teaching by both our students and the teaching assistants. Hence, if anything, we feel the impact of our changes would be even more in a "normal" year. 

The second factor that could shape the surveys is the change in instructor. We again believe that the expected effect here would be negative, as the previous instructor was very experienced and very well liked by both the students and the teaching assistants. The main instructor for the period relevant to this report was a junior faculty member with limited teaching experience. While this does not imply either a positive or negative effect, we find a negative effect on the outcomes of the survey more likely.

\section{Conclusion}
\label{sec:conclusion}
In this paper, we reported on our experience with changing the problem session structure. We observed that the new warm-up questions indeed bridged the gap between the lectures and the problem sessions, making them more easily accessible for students, as also commented on in the End of Semester Survey. While the grades did not change significantly, teaching assistants noted a significant improvement in student understanding, satisfaction, and confidence. Using the warm-up questions also made teaching easier for them, as they could address misconceptions the students had about the material separately from the problem solving aspects of the problem session. 

While this approach showed an improvement in student confidence, it did not lead directly to improved performance in the subject. We conjecture that this is due to the heterogeneous nature of the student cohort taking this subject and improving this will require investigating and addressing potential missing assumed knowledge. Extending our problem sessions further to cover such material is an option, though we need to be careful to ensure this does not negatively affect the students that already possess this knowledge by driving them away from the problem sessions in the early stages of the subject. 

Future work related to this report includes determining how to leverage the increased student confidence and engagement to also facilitate a significant increase in student performance and grades. We also aim to incorporate the changes reported on here in follow-up units, though we expect the impact of such changes there to be minor, as students already have an increased confidence in their algorithmic abilities from the unit reported on.

\paragraph{Acknowledgements}
We thank Alan Fekete for useful comments that helped shape this paper. We also thank our teaching assistants for their invaluable input on their experience teaching using the new problem sets.

\bibliographystyle{plain}
\bibliography{references}

\end{document}